\documentclass{article}

\usepackage{arxiv}

\usepackage[utf8]{inputenc} 
\usepackage[T1]{fontenc}    
\usepackage{hyperref}       
\usepackage{url}            
\usepackage{booktabs}       
\usepackage{amsfonts}       
\usepackage{nicefrac}       
\usepackage{microtype}      
\usepackage{lipsum}

\usepackage{graphicx}
\usepackage{caption}

\usepackage{amsfonts,amsmath,amssymb,amsthm}
\usepackage{bm,nicefrac}

\usepackage{algorithmicx}
\usepackage{algpseudocode}
\usepackage[]{algorithm2e}

\newtheorem{theorem}{Theorem}[section]

\title{An ``On The Fly'' Framework for Efficiently Generating Synthetic Big Data Sets}

\author{
  Karl Mason\\
  School of Electrical and Computer Engineering\\ 
  Georgia Institute of Technology\\ 
  Atlanta, GA, USA\\
  \texttt{kmason35@gatech.edu} \\
  \And
 Sadegh Vejdan \\
  School of Electrical and Computer Engineering\\ 
  Georgia Institute of Technology\\ 
  Atlanta, GA, USA\\
  \texttt{svejdan3@gatech.edu} \\
   \And
 Santiago Grijalva \\
  School of Electrical and Computer Engineering\\ 
  Georgia Institute of Technology\\ 
  Atlanta, GA, USA\\
  \texttt{sgrijalva@ece.gatech.edu} \\
}

\begin{document}
\maketitle

\begin{abstract}
Collecting, analyzing and gaining insight from large volumes of data is now the norm in an ever increasing number of industries. Data analytics techniques, such as machine learning, are powerful tools used to analyze these large volumes of data. Synthetic data sets are routinely relied upon to train and develop such data analytics methods for several reasons: to generate larger data sets than are available, to generate diverse data sets, to preserve anonymity in data sets with sensitive information, etc. Processing, transmitting and storing data is a key issue faced when handling large data sets. This paper presents an ``On the fly'' framework for generating big synthetic data sets, suitable for these data analytics methods, that is both computationally efficient and applicable to a diverse set of problems. An example application of the proposed framework is presented along with a mathematical analysis of its computational efficiency, demonstrating its effectiveness.  
\end{abstract}


\section{Introduction}
\label{sec:intro}



Many data analytics and machine learning methods require large volumes of data to effectively accomplish the tasks they are designed for. A well known example of this is image classification\cite{rawat2017deep}. In order to accurately classify images, a large database of images is required. Having access to such a large data set is problematic for many tasks, including face recognition. Due to this limitation, many researchers utilize synthetic data sets to develop machine learning algorithms. Generating synthetic data sets has the advantage of easily producing large and diverse data sets without the need for actually collecting the data. Of course it is imperative that the synthetic data set generated is an accurate representation of the actual problem. This typically requires additional computational effort to generate the data set, input from a domain expert to ensure that the synthetic data is accurate and original data sets.


The most common method used to generate synthetic data is to generate it in full and store entire synthetic data set in a tabular format \cite{mason2018meta}. This straightforward process is easy to implement and is appropriate for small data sets. This approach can require large amounts of storage space as the size of the data set increases. This can be alleviated to some extent by representing the data set in a graph structure for large relational databases \cite{park2014graph}, however this will still require vast amounts of storage.

Many learning tasks require such a vast amount of data to develop accurate models, that virtual environments are implemented generate the data to train these models. One of the best examples of this is self driving cars \cite{pan2017virtual}. In order to develop algorithms that can safely drive vehicles on public roads, a vast amount of training data is needed. This involves gathering large volumes of data, e.g. images data, to train the driving software. In order ease the development of these models, researchers in the field routinely use simulated virtual environments to help train the driving software. These simulated environments are effectively generating synthetic data representing what a vehicle would observe when driving on a public road. This simulated approach has the advantage of training the model for a larger number of scenarios than would be possible to collect in the real world. This approach relies heavily on the development of accurate simulators. Generative Adversarial Networks (GANs) are a recent development in machine learning that have made it possible to generate increasingly realistic synthetic images \cite{goodfellow2014generative}. GANs are highly desirable for generating synthetic image data sets for classification purposes due to the high quality and realism of the images they produce.

The approach of generating synthetic data in full before performing data analytics encounters a problem when the computational requirements exceed the available computational resources. In order to address this problem, an ``On the fly'' framework for synthetic data generation is proposed in this paper. This framework circumvents these large computational requirements by generating data in batches as needed. This framework requires less memory to store the current batch of data and any optional recovery information that is needed to regenerate some or all of the current data batch at a later date. This framework allows the user to generate data in batches during run time, the size of the batch can be selected based on the amount of RAM available. The framework is also general enough to be applicable to a range of problem domains.

The contributions of this paper are as follows:

\begin{enumerate}
    \item To propose a flexible ``On the fly'' framework for synthetic data generation that has less computational requirements than standard approaches and does not require a vast amount of storage.
    
    \item To present an example application of the proposed framework and demonstrate its applicability.
    
    \item To analyze the computational complexity of the proposed framework and demonstrate its applicability.
\end{enumerate}

The rest of the paper is structured as follows: Section 2 will give an overview of research relating to big data and synthetic data sets. Section 3 will outline the standard procedure of generating synthetic data sets and then present the proposed ``one the fly'' framework. A detailed discussion of the proposed framework and its computational requirements will be presented in Section 4. Finally, Section 5 will highlight what conclusions can be made as a result of the work presented in this paper.

\section{Background}
\label{sec:Background}


\subsection{Big Data and Machine Learning}
\label{sec:BigData}


Machine learning is one of the most widely used tools for data analytics in large data sets \cite{al2015efficient}. Machine learning methods are algorithms that learn from experience by interacting with a data set. Machine learning algorithms are used for a wide range of tasks relating to big data. One such task is that many data sets do not have labels. Clustering algorithms can be useful for such problems to discover groups in the data, e.g. clustering load profiles using smart meter data \cite{mcloughlin2015clustering} and clustering electricity price data \cite{vejdan2018expected}. Netflix is one of the most well known companies that heavily relies on machine learning methods to recommend shows to its users based on what it believes they will like \cite{gomez2016netflix}.

\subsection{Synthetic Data}
\label{sec:syntheticData}

When developing data analytics algorithms, it is not always possible to have complete data sets. This is particularly important for machine learning methods where the performance is heavily reliant on the data it is provided with. There are many reasons why an adequate data set might not be available. 

\begin{itemize}
    \item Privacy. The data that is required contains sensitive information and is therefore difficult to access.
    
    \item Collection Time. The data may take an impractically long time to collect.
    
    \item Cost. It may be too expensive collect the data or to gain access to it.
    
    \item Diversity. The data that is available is only representative of a small subset of possible scenarios and is therefore insufficiently diverse.
\end{itemize}

Many researchers generate synthetic data sets in order to overcome the issues outlined above \cite{ming2013bdgs}. The goal when generating synthetic data sets is to create a data set that is an accurate representation of what would be observed if real data were available. Aside from the accuracy of the data set, there are a number of challenges associated with generating large synthetic data relating to its computational cost. Generating large data sets requires a significant amount of computational resources, including: CPU time, RAM and disk storage. The amount of each of these resources required will vary depending on the size and nature of the data being generated.  

\subsection{Big Data in Industry}
\label{sec:BigDataIndustry}

There are many industries that face the challenges associated with handling large data sets. The finance sector is a prime example of one such industry \cite{fang2016big}. In high frequency trading, stock trading algorithms make trades in the order of milliseconds \cite{duhigg2009stock}. In order for these algorithms to successfully operate at such short timescales, very large volumes of data must be processed and analyzed in very short time frames. These data analytics methods are typically implemented for tasks such as stock price forecasting, assessing risk, portfolio selection, sentiment analysis, etc. The difficulty of these tasks is further increased by the format of the data that must be analyzed. For example, monitoring the news to assess the sentiment in the market would heavily rely on natural language processing to estimate the prevailing attitude in the market \cite{zhang2010trading}. The use of synthetic data is common practise for stock market data analylics \cite{ren2010recurrence}.

Healthcare is another industry where data analytics and machine learning tools are expected to heavily influence the field. Machine learning algorithms have been implemented to address a wide range of healthcare problems, from predicting the outbreak of diseases \cite{chen2017disease} to predicting patient mortality from sepsis \cite{taylor2016prediction}. One of the key issues surrounding the application of machine learning to these data sets is patient privacy. There are many regulations surrounding access of confidential patient information. The use of synthetic data sets is therefore very valuable to researchers as it can circumvent the issue of patient privacy, in addition to generating large and diverse data sets.


Another example is energy analytics. The power grid is currently undergoing significant changes with the deployment of Advanced Metering Infrastructure (AMI) and the collection of large volumes of data \cite{kezunovic2012fundamental}. It is hoped that these measurements will give utility providers the capabilities to better monitor and manage the grid \cite{kezunovic2013role}. Smart meters can make multiple measurements per second, which highlights the issue of scalability for large scale deployment in distribution feeders consisting of thousands of homes or in cities with millions of electricity customers \cite{hao2012smart}. When developing data analytics algorithms for smart meter analysis, researchers often rely on synthetic data sets \cite{liu2017smart}. This is due to the difficulty in obtaining adequate smart meter data sets for reasons such as privacy.

Technology companies are also faced with a wide range of problems that require effective analytics tools capable of handling large data sets. For example, Facebook utilized approximately 4.4 million labeled images of faces to train a neural network for the task of face recognition \cite{taigman2014deepface}. The problem of face recognition is a very challenging problem and therefore requires vast amounts of training data. Accumulating such a large data set involves a significant amount of time and effort.  As mentioned in the introduction, the development of Generative Adversarial Networks (GANs) is one of the key advancements in machine learning in recent years \cite{goodfellow2014generative}. GANs consist of two neural networks, one is used to generate an image and the other is to detect if the image is real or not. GANs have made it possible to generate synthetic images that are often indistinguishable from a real image. The effectiveness of GANs mean that they are promising candidates for generating large synthetic image data sets.

Many companies utilize cloud computing services to minimize their own in house computing infrastructure. Cloud services such as Microsoft azura are now widely used by companies for online data management. These online platforms require large data centers to power these online services. Such data centers must process data on the exabyte scale. Large amounts of energy is required to power such large data centers. Google has demonstrated that an energy saving of 40\% can be achieved by implementing machine learning to manage its data center \cite{gao2014machine}. Many researchers investigating data center management utilize synthetic data to represent workloads for their simulated data centers \cite{bahga2011synthetic} and for predicting cloud computing resources \cite{mason2018predicting}. This allows the data centers to be simulated in a wider range of scenarios than would otherwise be possible.

\section{Generating Synthetic Data}
\label{sec:GenData}





\begin{figure}[h]
    \centering
    \includegraphics[width=0.25\textwidth]{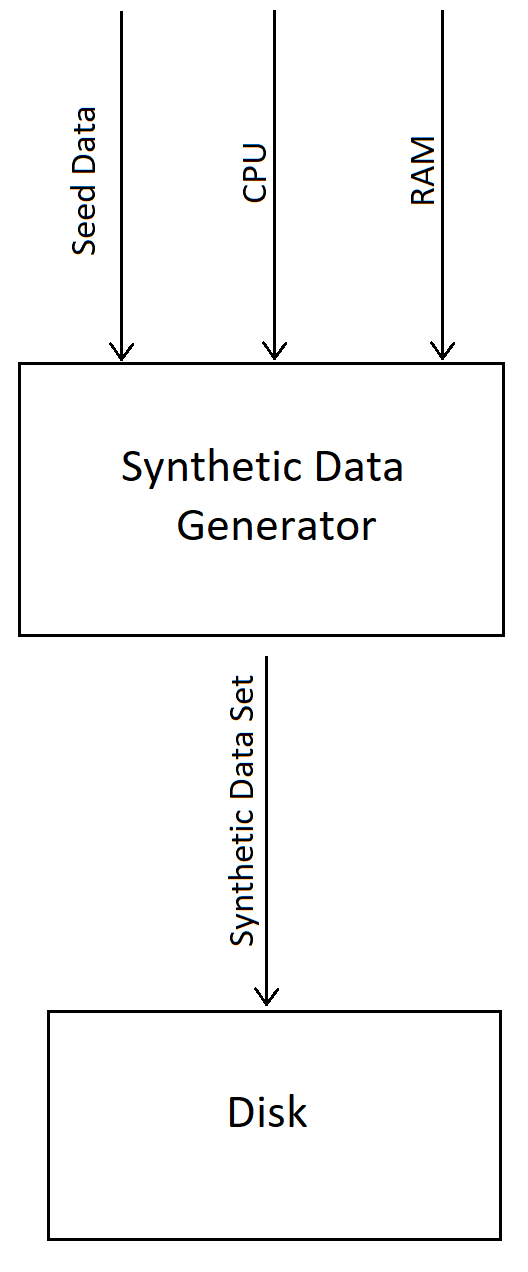}
    \caption{Standard Procedure for Synthetic Data Generation. 
    }
    \label{fig:SynthDataGen}
\end{figure}

As outlined in Section \ref{sec:Background}, there are many situations when generating synthetic data sets is necessary. The standard approach for doing this is to generate the synthetic data set in full before it is processed or analyzed. Figure \ref{fig:SynthDataGen} illustrates this process. It is common place for a relatively small amount of real world data to be used as a seed to generate larger data sets. The synthetic data generator generally consists of a number of processes and equations used to permutate the input seed data to generate new synthetic data. The generated data set is then written to disk all at once or in batches, depending on the size of the data set and computational constraints. This is a straightforward and intuitive approach to implement, and will work well when generating smaller data sets. 

\begin{figure*}[h]
    \centering
    \captionsetup{justification=centering}
 
    \includegraphics[width=0.7\textwidth]{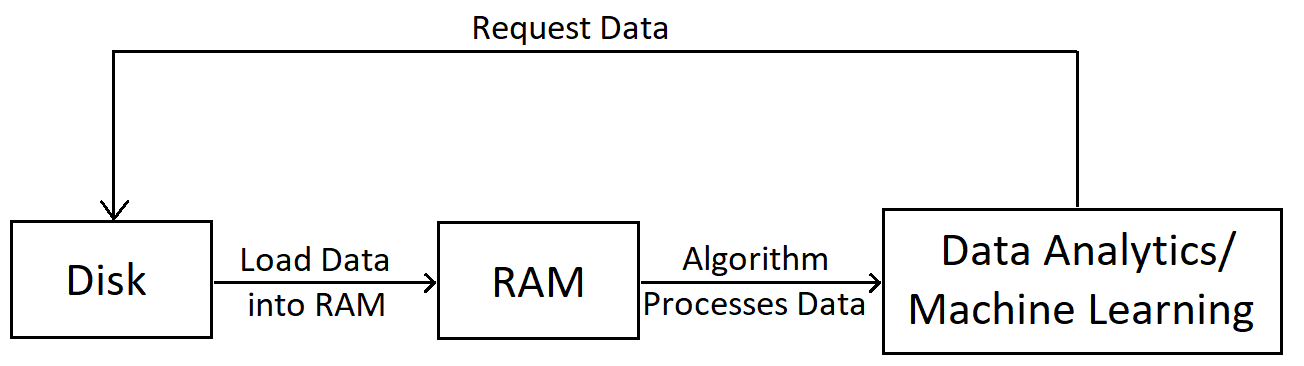}
    \caption{Data Analytics Flowchart. 
    }
    \label{fig:DataRead}
\end{figure*}

The drawback with this approach however is: 1) The large computational and time cost that must be paid upfront when generating large data sets. 2) The computational and time cost of reading the data into RAM from disk. When generating large data sets, a significant amount of CPU time and RAM must be spent upfront generating the data set which must be then transferred and stored on the disk. If limited computational resources are available (e.g. RAM and disk space on a desktop PC), the possible size of the synthetic data set is then severely limited by what the PC can produce in the time frame available. When utilizing the generated synthetic data set for performing analytics, the data must be read into RAM from the disk which takes a significant amount of time. This process is illustrated in Figure \ref{fig:DataRead}. The amount of data read into RAM is also limited by the RAM available. This will result in reading data from the disk in multiple batches in cases the amount of RAM is a small fraction of the data set size. This will further increase the demand on computation and time resources, as it will require transferring data from the disk multiple times. The cost of moving information into RAM from the disk when using real data sets can only be reduced by compressing the data or by upgrading the physical computing infrastructure. The upfront data generation cost and the cost of moving information to and from memory can be reduced when generating synthetic data sets by generating the data during execution time and keeping the batches of data in RAM. This is the key contribution of the proposed ''On the fly`` synthetic data generation framework, which scales better than the standard approach outlined here.

\subsection{''On the fly`` Generation}
\label{sec:OnTheFly}

This section will outline the proposed ''On the fly`` (OTF) framework for synthetic data generation. As mentioned in the previous section, the key advantage of the proposed OTF framework is the consolidation of the synthetic data generation phase and the data analytics phase from two distinct phases into one combined cycle. This has the advantage of minimizing the amount of time consuming data reading and writing to the disk. The proposed OTF framework achieves this by generating the synthetic data as it is required rather than all at once before it is needed. By doing this, the generated data is kept in RAM and is easily accessible by the algorithm. As the algorithm executes and requests the next batch of data, the current batch of data is discarded. By discarding the old batch, memory is freed up to generate the next batch of synthetic data. It should be noted here that when deleting the old batch of synthetic data, a small amount of computational resources are allocated to record the parameters used by the synthetic data generator to generate the old batch. This can be stored in RAM or written to disk depending on if they will likely be needed in the near future. The purpose of this is to be able to re-generate the old batch at later if desired. This step can be omitted if it is not required for a particular task.

Typically there are only a small number of parameters used to permutate seed data to generate synthetic data, e.g. scaling factors, weightings, etc. The exact number of parameters is problem specific, however it is almost always significantly smaller than the data set being generated. For example, in a scenario where energy consumption data is generated using a seed profile for a particular day in increments of 1 second. This seed data would correspond to a total of 86400 values (seconds in a day).

For simplicity, lets assume that the synthetic data generator generates data by scaling the seed profile and combining it with a scaled seed noise profile as outlined in Equation \ref{eqn:SynthEG}.

\begin{equation}
  \bm{D_g} = \bm{D_s} \times \lambda_1 + \bm{N_m} \times \lambda_2
  \label{eqn:SynthEG}
\end{equation}

Where $\bm{D_g}$ is the generated synthetic data, $\bm{D_s}$ is the s$^{th}$ seed data file, $\lambda_1$ and $\lambda_2$ are real valued constants $\in [0,1]$ and $\bm{N_m}$ is the m$^{th}$ noise profile. 

In order to reproduce the synthetic data $\bm{D_g}$ at a later date, 4 parameters need to be recorded $[s, \lambda_1, \lambda_2, m]$. This is 21600 times fewer parameters than the 86400 values that would need to be otherwise stored. This approach of storing the data generation parameters is therefore much more scalable than storing all of the generated data. In this example, if the data in $\bm{D_g}$ increases in length to 604800 (a week's worth of energy consumption at a 1 second resolution), the number of parameters stored by the proposed OTF framework does not increase. A significant amount of RAM and disk space would be required if all values in $\bm{D_g}$ were stored instead of just their generation parameters.

\subsection{Proposed Framework}
\label{sec:ProposedFrame}

\begin{figure*}[h]
    \centering
    \captionsetup{justification=centering}
 
    \includegraphics[width=0.7\textwidth]{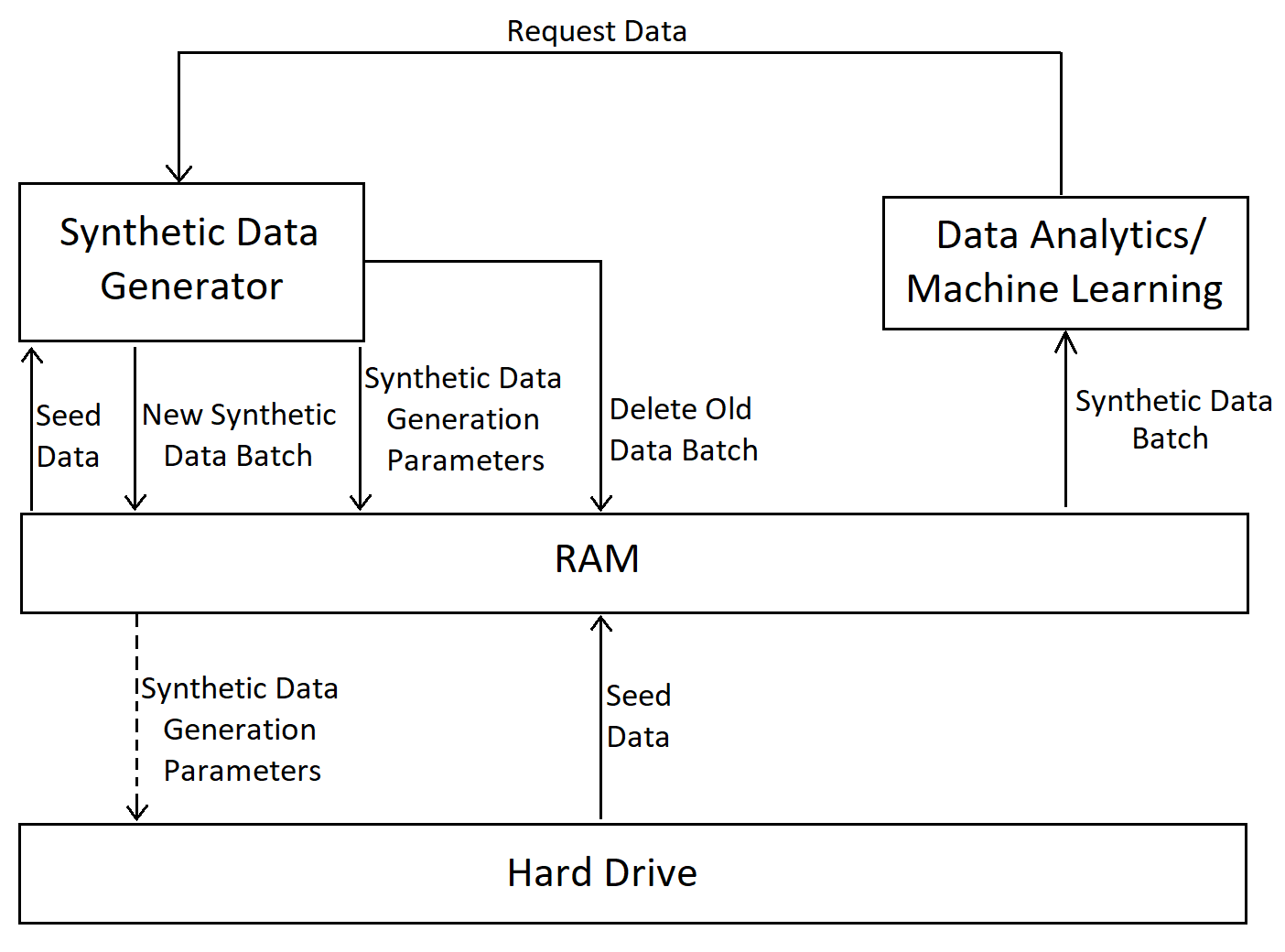}
    \caption{On The Fly Synthetic Data Generation Flowchart. 
    }
    \label{fig:OTFFlowchart}
\end{figure*}

The proposed structure of the OTF framework is outlined in Figure \ref{fig:OTFFlowchart}. In this framework, the data analytics algorithm makes a request from the synthetic data generator for a batch of generated data. The batch size is problem specific and also depends on the RAM available on the machine. This points to an advantage of the proposed OTF framework, data can be fed into RAM in manageable sizes depending on the resources available. This is important when considering large data sets which would be too large to store in memory. At this point the old data batch is deleted to free up memory. The data generation parameters are retained however, to regenerate the old batch if needed. The synthetic data generator then generates a new batch of data using the seed data that has been loaded into memory. The seed data is loaded into memory once and kept there to reduce the amount of time reading from the hard drive. Once the new batch of synthetic data is generated, it is kept in memory for the data analytics algorithm to access. Once the data analytics process is complete these data generation parameters are stored on the hard drive before the program terminates. The pseudo code in Algorithm \ref{Alg:OTF_Alg} describes the algorithm that governs the OTF framework presented in Figure \ref{fig:OTFFlowchart}.

\begin{algorithm}[]
\caption{``On The Fly'' Framework for Synthetic Data Generation Pseudocode}
\begin {algorithmic}
\State \textbf{Data Analytics Code:}
\State Create instance of Synthetic Data Generator = SDG\\
\While{Data Analytics NOT finished}{
    \State batch = SDG.requestData()
    \State perform data analytics on batch
}
\State SDG.SaveDataParameters()

\State\\
\State \textbf{Synthetic Data Generator Code:}
\State Load Seed Data into RAM
\Function{requestData}{\null}
  \State Remove old data batch
  \State Generate new batch of data using seed data
  \State Return new batch of data
 \EndFunction
\Function{SaveDataParameters}{\null}
  \State Write generated data parameters to file
 \EndFunction
\end{algorithmic}
\label{Alg:OTF_Alg}
\end{algorithm}

\subsection{Machine Learning Application Example}
\label{sec:MLAppEG}

Section \ref{sec:ProposedFrame} outlined a detailed overview of the proposed OTF framework and how it can applied in a general sense. This section will give a concrete example of how this framework can be applied to a specific task. The data set that will be utilized for this example will be the energy consumption classification problem, outlined in Section \ref{sec:OnTheFly}. The aim of this section is to demonstrate how the OTF framework can be implemented to generate synthetic consumption data on the fly, which is then used to train a neural network classifier to determine if the consumption profile for that day corresponds to a residential or commercial energy customer. Neural networks are a prominent area of machine learning research. These networks process an input signal, by a series of weighted connections and processing units with activation functions, to produce an output signal. For a more comprehensive description of neural networks, see the work of Schmidhuber et al. \cite{schmidhuber2015deep}.

In this example, the features that are input into the neural network are the energy consumption profiles. The aim of the neural network is to determine if a profile is representative of a residential (1) or commercial customer (0). These are the outputs of the network. In this example, there is only a limited amount of real world data available. This is the seed data that will be used by the synthetic data generator to generate larger data sets. We will use the same notation from Equation \ref{eqn:SynthEG} in Section \ref{sec:OnTheFly}. We have $s=10$ energy consumption profiles for 10 different customers each with a 1 second resolution and a time frame of a year (not a day as was the case in Section \ref{sec:OnTheFly}). This is amounts to $3.154e+7$ data points in each of the 10 seed data sets $\bm{D_s}$. We also have a total of m=20 noise data sets $\bm{N_m}$ (each of length $3.154e+7$) that will be used by the synthetic data generator to generate new data sets using Equation \ref{eqn:SynthEG}. The OTF framework can be applied to train the neural network to do this task as outlined in the pseudocode in Algorithm \ref{Alg:OTF_AlgApp}.

\begin{algorithm}[]
\caption{OTF Framework Implementation for Neural Network Classification of Residential and Commercial Customers}
\begin {algorithmic}
\State \textbf{Neural Network Code:}
\State Create neural network
\State Create instance of Synthetic Data Generator = SDG
\State Train and test data size = 50 yearly profiles
\State batchSize = 10 yearly profiles\\
\While{epoch e \textless maxEpoch}{
    \State sumError =0\\
    \While{batch t \textless numTrainBatches}{
        \State batch = SDG.requestData(batchSize)
        \State Split batch into 3650 daily consumption profiles\\
        \For{each day i in 3650}{
            \State Evaluate network on profile i
            \State Calculate error between network output and label
            \State sumError += error}
        \State t++
    }
    \State backpropagate sumError and update weights
    \State e++
    }\\
\While{batch t \textless numTestBatches}{
        \State batch = SDG.requestData(batchSize)
        \State Split batch into 3650 daily consumption profiles\\
        \For{each day i in 3650}{
            \State Evaluate network on profile i
            \State Make prediction for profile i
        }
        \State t++
}
\State SDG.SaveDataParameters()
\State
\State \textbf{Synthetic Data Generator Code:}
\State Load Seed Data into RAM [$\bm{D_s}, \bm{N_m}$]
\Function{requestData}{batchSize}
  \State Store old data generation parameters [$s, m, \lambda_1, \lambda_2$]
  \State Remove old data batch
  \State Generate batch $\bm{D_g}$ (Equation \ref{eqn:SynthEG})
  \State Return $\bm{D_g}$
 \EndFunction
 \Function{SaveDataParameters}{\null}
  \State Write generated data parameters to file
 \EndFunction
\end{algorithmic}
\label{Alg:OTF_AlgApp}
\end{algorithm}

As can be seen from this example. A neural network classifier can be trained and tested using data generated using the OTF framework. This framework can generate batches of data of a size the user selects and train the neural network without the heavy computational overhead of generating very large data sets and then loading these data sets to and from the hard drive. The OTF framework is suited for situations where limited computational resources are available. It should be noted here that if the synthetic data set is needed more than once, the proposed framework would need to generate the data more than once. This would take additional time, the length of which would vary depending on the nature of the problem. This would be offset to some degree by the fact that OTF does not require multiple instances of data reading and writing to the hard drive which is a time consuming task. The main advantage of the proposed OTF approach is its flexibility. OTF can generate big data sets with low RAM, disk and time requirements which is important as computational resources are often limited. The other key advantage is the ease at which the synthetic data generation process can be modified. For instance, Equation \ref{eqn:SynthEG} can be modified to read in new seed files when generating synthetic date with almost no effort. If the synthetic data set were to be generated in full upfront, making any changes to how it was generated would require re-generating the full data set. This would involve many instances of time consuming read and write commands to the disk, as discussed earlier. As can be seen in Figure \ref{fig:OTFFlowchart}, the only read/write steps involve the synthetic data generation parameters and reading the seed data into memory. Each of these are carried out just once. In contrast, Figure \ref{fig:DataRead} illustrates that information must be read from the disk drive multple times if the OTF framework is not implemented.






    

\section{Discussion}
\label{sec:Discussion}

\subsection{Analysis of Computational Cost}
\label{sec:CompAnalysis}

Lets first consider a comparison between the computational resources required for generating a data set in batches and generating the data all at once.

The relationship between the data set size \textit{D} and the batch size \textit{B} is calculated as outlined in Equation \ref{eqn:BatchSize}.

\begin{equation}
  N_B B = D 
  \label{eqn:BatchSize}
\end{equation}

Where \textit{$N_B$} is the number of batches. The relationship between the RAM required for the entire data set \textit{$RAM_{D}$} and for a batch of data \textit{$RAM_{B}$} is calculated using Equation \ref{eqn:BatchRAM}.

\begin{equation}
  RAM_B = \frac{RAM_D}{N_B} 
  \label{eqn:BatchRAM}
\end{equation}

This inverse relationship between \textit{$RAM_{B}$} and \textit{$N_B$} demonstrates that unless a large amount of RAM is available, the data will need to be read in batches for large data sets. The number of batches depending on the size of the data set and the RAM available.

Next we can compare the amount of computational resources required for pre-generating the data set before the data analytics step versus generating it on the fly. Consider first the amount of RAM required to read a batch of data from a file which is simply \textit{$RAM_B$}. The OTF framework RAM requirements \textit{$RAM_{OTF}$} per batch are calculated using Equation \ref{eqn:BatchRAM_OTF}.
\begin{equation}
  RAM_{OTF} = RAM_B + i * RAM_{P}
  \label{eqn:BatchRAM_OTF}
\end{equation}

Where \textit{i} is the current batch number and \textit{$RAM_{P}$} is the RAM required to hold the data generation parameters of a given batch discussed in Section \ref{sec:OnTheFly} (Equation \ref{eqn:SynthEG}). As previously discussed, the number of these parameters is orders of magnitude smaller than the number of parameters in the batch of data, therefore \textit{$RAM_{P}$} is significantly smaller than \textit{$RAM_{B}$}. The recording of these parameters is also optional, in cases where these parameters are not needed for future use \textit{$RAM_{OTF} = RAM_B$}. Therefore the RAM requirements of the proposed OTF framework are approximately equivalent to reading from a pre-generated synthetic data set. 

When comparing the pre-generation approach to the OTF framework, it is also important to consider hard drive requirements. The amount of disk space required for the pre-generation approach $Disk_{PG}$ is calculated using Equation \ref{eqn:PreGenDisk}.
\begin{equation}
  Disk_{PG} = Disk_{Seed} + Disk_{D}
  \label{eqn:PreGenDisk}
\end{equation}

Where \textit{$Disk_{Seed}$} is the disk requirements to hold the seed data and \textit{$Disk_{D}$} is the disk requirements to hold the generated synthetic data set. In contrast the disk requirements of the proposed OTF framework \textit{$Disk_{OTF}$} are calculated using Equation \ref{eqn:OTFDisk}, where \textit{$Disk_{P}$} is the disk space required to hold the data generation parameters for a given batch. 
\begin{equation}
  Disk_{OTF} = Disk_{Seed} + N_{B} * Disk_{P}
  \label{eqn:OTFDisk}
\end{equation}

We can now prove that OTF requires much less hard drive space than the standard data pre-generation approach. Let us first define $\Bar{P} = N_B * P$ as the total number of parameters needed by the OTF framework to regenerate the full data set at a later date.


\begin{theorem}[OTF requires less disk space than Pre-Generating the data set]

\label{theorem:OTF_DiskSpace}
When generating synthetic data, $Disk_{OTF} < Disk_{PG}$.

\begin{proof}
From Equation \ref{eqn:PreGenDisk}
\[Disk_{PG} = Disk_{Seed} + Disk_{D}\]

From Equation \ref{eqn:OTFDisk}
\[Disk_{OTF} = Disk_{Seed} + N_{B} * Disk_{P}\]

$Disk_{PG} - Disk_{OTF}$ gives
\[Disk_{D} - N_{B} * Disk_{P}\]

Which is equivalent to
\[Disk_{D} - Disk_{\Bar{P}}\]

Assuming 
\[D > \Bar{P}\]

It follows that
\[Disk_{D} - Disk_{\Bar{P}} > 0\]

Which implies

\[Disk_{OTF} < Disk_{PG}\]


\end{proof}

\end{theorem}

It can be seen here that OTF does not need to store the generated data set, instead it stores the parameters required to generate the data set. For large data sets, the disk required to hold these generation parameters is orders of magnitude lower than the disk requirements to hold the full data set. This low disk space requirement is a key advantage of the proposed OTF framework. 


Next we can compare the performance of the proposed OTF framework versus the standard data pre-generation in terms of the number of times data must be read/written to the disk. Let us first consider the number of disk read write instances only during the data analytics process using the pre-generated data. This is simply \textit{$N_B$}. However, we must also consider the number of disk read/writes during the data generation process to make a fair comparison. The total number of disk read/write instances \textit{$N_{RW}^{PG}$} during data generation and the data analytics process is calculated using Equation \ref{eqn:PreGenRW} for the data pre-generation approach.
\begin{equation}
  N_{RW}^{PG} = N_{S} +  2 * N_{B}
  \label{eqn:PreGenRW}
\end{equation}
Where \textit{$N_{S}$} is the number of seed data files that must be loaded into RAM during data generation. The generated data must then be written to disk \textit{$ N_{B}$} times once it is generated. The second \textit{$ N_{B}$} corresponds to the reading of the data from the disk when the data analytics algorithm needs it. It should be noted here that it is assumed that the same batch size is used to write data to the disk once generated as the batch size used to read the data during the analytics process. In contrast, the number of disk read/writes for the proposed OTF framework is calculated using Equation \ref{eqn:OTF_RW}.
\begin{equation}
  N_{RW}^{OTF} = N_{S} + 1
  \label{eqn:OTF_RW}
\end{equation}

\begin{theorem}[One full cycle of data generation and analytics requires fewer disk read/write operations using OTF than by Pre-Generating the data set]

\label{theorem:OTF_DiskRW}
For one full cycle of data generation and analysis, $N_{RW}^{OTF} < N_{RW}^{PG}$ where $N_{RW}^{OTF}$ is the number of OTF disk read/write instances and $N_{RW}^{PG}$ is the number of disk read/write instances of the data pre-generation approach.

\begin{proof}
From Equation \ref{eqn:PreGenRW}
\[N_{RW}^{PG} = N_{S} +  2 * N_{B}\]

From Equation \ref{eqn:OTF_RW}
\[N_{RW}^{OTF} = N_{S} + 1\]

$N_{RW}^{PG} - N_{RW}^{OTF}$ gives
\[2 * N_{B} - 1\]

Assuming 
\[B > 1 \]

It follows that
\[2 * N_{B} - 1 > 0\]

Which implies

\[N_{RW}^{OTF} < N_{RW}^{PG}\]


\end{proof}

\end{theorem}

The ``1'' read/write corresponds to writing the data generation parameters to a file for future use in the OTF framework. This is an optional step but it is likely needed in many applications. By contrasting \textit{$N_{RW}$} for both the pre-generation approach and the proposed OTF, it is clear the number of disk read/writes is far fewer. Reading and writing from the disk is a time consuming process so this is an important advantage of the proposed OTF framework.

Finally, we can compare the time taken for one full cycle of generating data and passing the generated data to the data analytics module in batches for both the pre-generation approach and the proposed OTF approach. We will ignore the time taken for whatever data analytics procedure is being conducted since it is the same for both processes. Let us first define \textit{$T_{R,S}$} as the time taken to read in the seed data from disk, \textit{$ T_{GB} $} as the time taken to generate a batch of synthetic data, \textit{$ T_{W,B} $} as the time taken to write a batch of synthetic data to disk, \textit{$ T_{R,B}$} as the time taken to read a batch of data into memory, \textit{$T_{W,P}$} as the time taken to write the data generation parameters to the disk for a given batch of size \textit{$B$}, \textit{$\Bar{P}$} as the data generation parameters for the full data set, \textit{$D$} as the complete data set and \textit{$N_B$} as the number of batches. For the pre-generation approach the total time \textit{$T_{PG}$} is calculated using Equation \ref{eqn:PreGenTime}.
\begin{equation}
  T_{PG}= T_{R,S} + N_{B} * T_{GB} + N_{B} * T_{W,B} + N_{B} * T_{R,B}
  \label{eqn:PreGenTime}
\end{equation}
The total time taken to generate and pass the generated data to the data analytics module for the proposed OTF framework \textit{$T_{OTF}$} is calculated using Equation \ref{eqn:OTFTime}.
\begin{equation}
  T_{OTF}= T_{R,S} + N_{B} * T_{GB} + T_{W,\Bar{P}}
  \label{eqn:OTFTime}
\end{equation}

\begin{theorem}[One full cycle of data generation and analytics takes less computational time using OTF than Pre-Generating the data set]

\label{theorem:OTF_ComputationalTime}
For one full cycle of data generation and analysis, $T_{OTF} < T_{PG}$ where $T_{OTF}$ is the OTF computational time and $T_{PG}$ is the computational time of the data pre-generation approach.

\begin{proof}
From Equation \ref{eqn:PreGenTime}
\[T_{PG} = T_{R,S} + N_{B} * T_{GB} + N_{B} * T_{W,B} + N_{B} * T_{R,B}\]

From Equation \ref{eqn:OTFTime}
\[T_{OTF}= T_{R,S} + N_{B} * T_{GB} + T_{W,\Bar{P}}\]

$T_{PG} - T_{OTF}$ gives
\[ N_{B} * T_{W,B} + N_{B} * T_{R,B} - T_{W,\Bar{P}}\]

Assuming 
\[\Bar{P} < D \]

It follows that
\[T_{W,\Bar{P}} < T_{W,D} \]

Using Equation \ref{eqn:BatchSize}
\[T_{W,\Bar{P}} < N_{B} * T_{W,B} \]

It is therefore true that
\[ N_{B} * T_{W,B} + N_{B} * T{R,B} - T_{W,\Bar{P}} > 0\]

Which implies

\[T_{OTF} < T_{PG} \]


\end{proof}

\end{theorem}
Let us now consider a numerical example to evaluate the time taken for one full cycle of the pre-generation vs the OTF framework. Lets assume that in this example, the seed data size $S = 10$ GB and the desired synthetic data set size $D = 100$ GB, to be created in $N_B = 20$ batches of size $B = 5$ GB. We will consider a hard disk drive (HDD) and assume the read time is the same as the write time, i.e. $T_{R,1GB} = T_{W,1GB} = 10$ sec. Therefore the time taken to read and write a batch of data $T_{R,B} = T_{W,B} = 50$ sec. This is a reasonable time for a HDD. Next, we must estimate the time taken to generate a batch of synthetic data $T_{G,B}$. This will vary significantly depending on the nature of the data, however in this case lets assume $T_{G,B} = 5$ sec. The last parameter that must be estimated is the time taken to write the data generation parameters for the full synthetic data set $\Bar{P}$. The example in Section \ref{sec:OnTheFly} estimated $\Bar{P}$ to be $21600$ times smaller than the data set $D$. We can then estimate the size of $\Bar{P} \approx 0.0046$ GB for $D = 100$ GB. Finally, these values can be inserted into Equations \ref{eqn:PreGenTime} and \ref{eqn:OTFTime} to estimate the time taken for one full cycle of data generation for pre-generation and OTF respectively to be $T_{PG} = 2,300$ sec and $T_{OTF} = 300$ sec. In this example, OTF requires a time $T_{OTF} = 0.13$ $T_{PG}$. It is therefore reasonable to conclude that for one full cycle of data generation, the proposed OTF framework will require significantly less computational time than the standard pre-generation process.

\subsection{Summary}
\label{sec:summary}

The previous section demonstrated the many advantages of the proposed OTF framework. The OTF framework requires only a small fraction of additional RAM the reading from a pre-generated synthetic data set approach. This is to record the data generation parameters in order to recreate the data at a later date if needed. If this is not needed, its RAM requirements are the same. In terms of hard drive requirements, it is proved that the proposed framework requires less disk space than the standard approach. The previous section also proved that OTF has a fewer number of disk read/write instances and a shorter execution time for one cycle of generating a full data set and performing data analytics. OTF is also flexible by easily allowing the modification of the synthetic data generation process. It is a scalable framework to generate larger data sets and is applicable to a wide range problem domains, data sets and data analytics methods. Often when conducting research, it is not clear what the synthetic data set should look like. This means that multiple scenarios must be evaluated, which corresponds to creating \textit{N} different synthetic data sets with different characteristics. The example at the end of Section \ref{sec:CompAnalysis} demonstrated that OTF can provide a data generation time saving of $33.3$ minutes when compared to pre-generating the data. This saving would be increased to $N \times 33.3$ minutes when considering multiple scenarios.

The OTF framework is proposed for problems involving large data sets. If the data set is small and the data will be needed for many future instances, OTF will involve recreating the same data set multiple times. The extent to which this will take extra computational time is not clear, since the OTF framework has significantly fewer instances of reading and writing data to the disk, which is a time consuming task. For these cases with smaller data sets with a large amount of, it is advised to used the synthetic data pre-generation approach as the amount of computation required to process the data is not a problem. The OTF framework is proposed to handle larger data sets. OTF also requires a small amount of additional effort to implement, e.g. recording data generation parameters for future use. This is not a significant amount of additional effort however, and is outweighed by the advantages outlined above.

\subsection{Impact}
\label{sec:impact}

There are many advantages of the proposed OTF framework as outlined in the previous section. It is hoped that the strengths of OTF synthetic data generation can benefit researchers in a broad range of domains and disciplines. There are multiple problems that OTF can have a positive impact, including generating: stock market data \cite{ren2010recurrence}, patient data \cite{guan2018generation}, electrical load data \cite{liu2017smart}, cloud data \cite{bahga2011synthetic}, etc. These are just a small sample of the problems that the OTF framework is applicable to.

It is also hoped that the OTF framework can aid computer science researchers develop more powerful and effective data analytics algorithms for large data sets, particularly in the field of machine learning. As was demonstrated in Section \ref{sec:MLAppEG}, OTF is well suited to generating synthetic data for neural networks. OTF would be equally well suited to other machine learning methods, e.g. convolutional neural networks \cite{rawat2017deep}, linear regression \cite{seber2012linear}, k-means \cite{jain2010data}, etc. OTF also is suitable for a wide range of problem types including: classification, regression, clustering, etc. Another significant impact of the proposed OTF framework is its potential to benefit researchers with limited computational resources and enable them generate and analyze large data sets.

\section{Conclusion}
\label{sec:Conclusion}

This paper proposes a new ``On the fly'' framework for generating large synthetic data sets. Synthetic data sets are necessary for performing data analytics in a wide range of problem areas. The proposed OTF framework is a computationally efficient and scalable method for generating large synthetic data sets that is broadly applicable to many problem types. The contributions of this research are:

\begin{itemize}
    \item To propose a novel method of synthetic data generation for large data sets.
    \item To demonstrate how OTF would be applied to classify residential and commercial electricity customers.
    \item To prove the efficiency of the proposed OTF framework in terms of disk space, number of read write instances to the disk and execution time.
\end{itemize}

\bibliographystyle{unsrt}  


\begin{thebibliography}{10}

\bibitem{rawat2017deep}
Waseem Rawat and Zenghui Wang.
\newblock Deep convolutional neural networks for image classification: A
  comprehensive review.
\newblock {\em Neural computation}, 29(9):2352--2449, 2017.

\bibitem{mason2018meta}
Karl Mason, Jim Duggan, and Enda Howley.
\newblock A meta optimisation analysis of particle swarm optimisation velocity
  update equations for watershed management learning.
\newblock {\em Applied Soft Computing}, 62:148--161, 2018.

\bibitem{park2014graph}
Yubin Park, Mallikarjun Shankar, Byung-Hoon Park, and Joydeep Ghosh.
\newblock Graph databases for large-scale healthcare systems: A framework for
  efficient data management and data services.
\newblock In {\em Data Engineering Workshops (ICDEW), 2014 IEEE 30th
  International Conference on}, pages 12--19. IEEE, 2014.

\bibitem{pan2017virtual}
Xinlei Pan, Yurong You, Ziyan Wang, and Cewu Lu.
\newblock Virtual to real reinforcement learning for autonomous driving.
\newblock {\em arXiv preprint arXiv:1704.03952}, 2017.

\bibitem{goodfellow2014generative}
Ian Goodfellow, Jean Pouget-Abadie, Mehdi Mirza, Bing Xu, David Warde-Farley,
  Sherjil Ozair, Aaron Courville, and Yoshua Bengio.
\newblock Generative adversarial nets.
\newblock In {\em Advances in neural information processing systems}, pages
  2672--2680, 2014.

\bibitem{al2015efficient}
Omar~Y Al-Jarrah, Paul~D Yoo, Sami Muhaidat, George~K Karagiannidis, and Kamal
  Taha.
\newblock Efficient machine learning for big data: A review.
\newblock {\em Big Data Research}, 2(3):87--93, 2015.

\bibitem{mcloughlin2015clustering}
Fintan McLoughlin, Aidan Duffy, and Michael Conlon.
\newblock A clustering approach to domestic electricity load profile
  characterisation using smart metering data.
\newblock {\em Applied energy}, 141:190--199, 2015.

\bibitem{vejdan2018expected}
Sadegh Vejdan and Santiago Grijalva.
\newblock The expected revenue of energy storage from energy arbitrage service
  based on the statistics of realistic market data.
\newblock In {\em 2018 IEEE Texas Power and Energy Conference (TPEC)}, pages
  1--6. IEEE, 2018.

\bibitem{gomez2016netflix}
Carlos~A Gomez-Uribe and Neil Hunt.
\newblock The netflix recommender system: Algorithms, business value, and
  innovation.
\newblock {\em ACM Transactions on Management Information Systems (TMIS)},
  6(4):13, 2016.

\bibitem{ming2013bdgs}
Zijian Ming, Chunjie Luo, Wanling Gao, Rui Han, Qiang Yang, Lei Wang, and
  Jianfeng Zhan.
\newblock Bdgs: A scalable big data generator suite in big data benchmarking.
\newblock In {\em Workshop on Big Data Benchmarks}, pages 138--154. Springer,
  2013.

\bibitem{fang2016big}
Bin Fang and Peng Zhang.
\newblock Big data in finance.
\newblock In {\em Big data concepts, theories, and applications}, pages
  391--412. Springer, 2016.

\bibitem{duhigg2009stock}
Charles Duhigg.
\newblock Stock traders find speed pays, in milliseconds.
\newblock {\em Sat}, 1:05, 2009.

\bibitem{zhang2010trading}
Wenbin Zhang and Steven Skiena.
\newblock Trading strategies to exploit blog and news sentiment.
\newblock In {\em Fourth International AAAI Conference on Weblogs and Social
  Media}, 2010.

\bibitem{ren2010recurrence}
Fei Ren, Wei-Xing Zhou, et~al.
\newblock Recurrence interval analysis of trading volumes.
\newblock {\em Physical Review E}, 81(6):066107, 2010.

\bibitem{chen2017disease}
Min Chen, Yixue Hao, Kai Hwang, Lu~Wang, and Lin Wang.
\newblock Disease prediction by machine learning over big data from healthcare
  communities.
\newblock {\em Ieee Access}, 5:8869--8879, 2017.

\bibitem{taylor2016prediction}
R~Andrew Taylor, Joseph~R Pare, Arjun~K Venkatesh, Hani Mowafi, Edward~R
  Melnick, William Fleischman, and M~Kennedy Hall.
\newblock Prediction of in-hospital mortality in emergency department patients
  with sepsis: A local big data--driven, machine learning approach.
\newblock {\em Academic emergency medicine}, 23(3):269--278, 2016.

\bibitem{kezunovic2012fundamental}
Mladen Kezunovic, Santiago Grijalva, Papiya Dutta, and Ayusman Roy.
\newblock The fundamental concept of unified generalized model and data
  representation for new applications in the future grid.
\newblock In {\em 2012 45th Hawaii International Conference on System
  Sciences}, pages 2096--2103. IEEE, 2012.

\bibitem{kezunovic2013role}
Mladen Kezunovic, Le~Xie, and Santiago Grijalva.
\newblock The role of big data in improving power system operation and
  protection.
\newblock In {\em 2013 IREP Symposium Bulk Power System Dynamics and Control-IX
  Optimization, Security and Control of the Emerging Power Grid}, pages 1--9.
  IEEE, 2013.

\bibitem{hao2012smart}
Xiaohong Hao, Yongcai Wang, Chenye Wu, Amy~Yuexuan Wang, Lei Song, Changjian
  Hu, and Lu~Yu.
\newblock Smart meter deployment optimization for efficient electrical
  appliance state monitoring.
\newblock In {\em 2012 IEEE Third International Conference on Smart Grid
  Communications (SmartGridComm)}, pages 25--30. IEEE, 2012.

\bibitem{liu2017smart}
Xiufeng Liu, Lukasz Golab, Wojciech Golab, Ihab~F Ilyas, and Shichao Jin.
\newblock Smart meter data analytics: systems, algorithms, and benchmarking.
\newblock {\em ACM Transactions on Database Systems (TODS)}, 42(1):2, 2017.

\bibitem{taigman2014deepface}
Yaniv Taigman, Ming Yang, Marc'Aurelio Ranzato, and Lior Wolf.
\newblock Deepface: Closing the gap to human-level performance in face
  verification.
\newblock In {\em Proceedings of the IEEE conference on computer vision and
  pattern recognition}, pages 1701--1708, 2014.

\bibitem{gao2014machine}
Jim Gao.
\newblock Machine learning applications for data center optimization.
\newblock 2014.

\bibitem{bahga2011synthetic}
Arshdeep Bahga and Vijay~Krishna Madisetti.
\newblock Synthetic workload generation for cloud computing applications.
\newblock {\em Journal of Software Engineering and Applications}, 4(07):396,
  2011.

\bibitem{mason2018predicting}
Karl Mason, Martin Duggan, Enda Barrett, Jim Duggan, and Enda Howley.
\newblock Predicting host cpu utilization in the cloud using evolutionary
  neural networks.
\newblock {\em Future Generation Computer Systems}, 86:162--173, 2018.

\bibitem{schmidhuber2015deep}
J{\"u}rgen Schmidhuber.
\newblock Deep learning in neural networks: An overview.
\newblock {\em Neural networks}, 61:85--117, 2015.

\bibitem{guan2018generation}
Jiaqi Guan, Runzhe Li, Sheng Yu, and Xuegong Zhang.
\newblock Generation of synthetic electronic medical record text.
\newblock In {\em 2018 IEEE International Conference on Bioinformatics and
  Biomedicine (BIBM)}, pages 374--380. IEEE, 2018.

\bibitem{seber2012linear}
George~AF Seber and Alan~J Lee.
\newblock {\em Linear regression analysis}, volume 329.
\newblock John Wiley \& Sons, 2012.

\bibitem{jain2010data}
Anil~K Jain.
\newblock Data clustering: 50 years beyond k-means.
\newblock {\em Pattern recognition letters}, 31(8):651--666, 2010.

\end{thebibliography}





\end{document}